\documentclass[12pt,a4paper]{amsart}
\usepackage{amssymb}
\usepackage[final]{showkeys}
\usepackage{microtype}
\usepackage[utf8]{inputenc}
\usepackage{amsmath,amssymb}
\usepackage{cite}
\usepackage{nameref,hyperref}
\usepackage{graphicx}
\usepackage[numbers]{natbib}

\def\e{\rm {e}}
\def\G{\mathcal{G}}
\def\H{\mathcal{H}}

\allowdisplaybreaks

\begin{document}
\title[]{Time Fractional Cable Equation and Applications in Neurophysiology}

 \author{Silvia Vitali$^1$}
     \address{${}^1$Dipartimento di Fisica e Astronomia (DIFA), 
    	 University of  Bologna ``Alma Mater Studiorum", Italy}
    \email{silvia.vitali4@unibo.it \  (Corresponding Author)}
    
    \author{Gastone Castellani$^2$}
   \address{${}^2$Dipartimento di Fisica e Astronomia (DIFA), 
    	 University of  Bologna ``Alma Mater Studiorum", Italy} 
    	 \email{gastone.castellani@unibo.it}

    \author{Francesco Mainardi$^3$}
    	 \address{${}^3$Dipartimento di Fisica e Astronomia (DIFA), 
    	 University of  Bologna ``Alma Mater Studiorum", and INFN, Italy}
	 \email{francesco.mainardi@bo.infn.it }


    \date{\today}





\begin{abstract}
We propose an extension of the cable equation by introducing a Caputo time fractional derivative. 
The fundamental solutions of the most common boundary problems are 
derived analitically via Laplace Transform, and result be written in terms of known special functions. 
This generalization could be useful to describe anomalous diffusion phenomena with leakage as signal conduction in spiny dendrites.
The presented solutions are computed in Matlab and plotted.
\\     
    {\bf   
     Chaos, Solitons and Fractals (2017):
Special Issue on Future Directions in Fractional Calculus.
    Guest Editors: Mark M. Meerschaert, Bruce J. West, Yong Zhou.
    Published on line 29 April 2017;  
    \\
    http://dx.doi.org/10.1016/j.chaos.2017.04.043
}
\end{abstract}

\maketitle

\section*{Introduction}

The one dimensional cable model is treated in neurophysiology to model the electrical conduction of non-isopotential excitable cells,
we remind as example to the textbooks of Johnston and Wu (1994) \cite{Johnston}, Weiss (1996) \cite{Weiss} and Tuckwell (1988) \cite{Tukwell},
and for some mathematical details to Magin (2006) \cite{Magin}. 
In particular it describes the spatial 
and the temporal dependence of transmembrane potential $V_m(x,t)$ along the
axial $x$ direction of a cylindrical nerve cell segment.
The membrane behaviour is summarized by an electrical circuit with an axial internal resistance $r_i$, and a transmembrane capacitance $c_m$ and 
a transmembrane resistance $r_m$ 
in parallel, connecting
the inner part to the outside \cite{Johnston}. External axial resistance could be eventually included. Transmembrane potential is generated
by ionic concentration gradient across the membrane, and is maintained non null at rest (no current) by a combination of passive and active cell mechanisms.
Equivalent models 
can in fact be derived from the Nernst-Planck equation for 
electro-diffusive motion of ions , see Qian and Sejnowski (1989)\cite{Qian}.

Cell excitation can be caused by electro stimulation of the membrane. The consequent variation in transmembrane potential is transmitted along the cell segment.
The resulting differential equation for the transmembrane potential 
takes the form of a standard diffusion equation with an extra term to account leakage of ions out of the membrane, it results
in a decay of the electric signal in space and in time:
\begin{equation}\label{linear}
 \lambda^2\frac{\partial^2V_m(x,t)}{\partial x^2}-\kappa \frac{\partial V_m(x,t)}{\partial t}-V_m(x,t)=0\,,
\end{equation}
$\lambda=\sqrt{r_m/r_i}$ and $\kappa=r_m c_m$ are the characteristic space and time scales of the process,
determined by the values of the membrane resistance and capacitance per unit length of the system, 
see e.g. \cite{Magin}.
For simplicity in the rest of this work, following \cite{Magin},
we will use the dimensionless scaled variables $X=x/\lambda$ and $T=t/\kappa$,  
so that we consider the equation 
\begin{equation}\label{linear-scaled}
 \frac{\partial^2V_m(X,T)}{\partial X^2}-
  \frac{\partial V_m(X,T)}{\partial T}-V_m(X,T)=0\,.
\end{equation}

Some interesting quantities to neurophysiology are connected to First kind boundary condition (the Signalling Problem) and Second Kind boundary condition problems.
Signalling Problem is interesting to understand how the system evolves when excited at one
end with a specific potential profile, Second Kind boundary condition problem is interesting because it can be related to the profile of a current injected across the membrane.  

In Signalling Problems the cable  is considered of semi-infinite length
($0\le X<\infty$), initially quiescent for $T<0$ and excited for $T\ge 0$ at the accessible end ($X=0$) with a given input in  membrane potential $V_m(0,T) = g(T)$.   
The solution can be derived via the Laplace Transform (LT) approach:
\begin{equation}
  \frac{\partial^2V_m(X,T)}{\partial X^2}=
  (s+1)V_m(X,T)\,,
\end{equation}
 and the LT of the solution results 
\begin{equation}
 \widetilde{V}_m(X,s)=g(s)e^{-\sqrt{s+1}X}\,.
\end{equation}

Relevant cases are impulsive input $g(T)= \delta(T)$
and unit step input  $g(T) = \theta(T)$ where $\delta(T)$  and $\theta(T)$ denote the Dirac  and the Heaviside functions, respectively. The  solutions
corresponding to these inputs can be obtained by LT inversion \cite{Magin} and read in our notation
\begin{equation}\label{linsol}
 \G_{s}(X,T)=\frac{X}{\sqrt{4\pi T^3}}\e^{-(\frac{X^2}{4T}+T)}\,,
\end{equation}
 and 
\begin{equation}\label{linsol-step}
 \H_{s}(X,T)=\int_0^T \G_{s}(X,T')\, dT'\,.
\end{equation}
We  refer to $\G_s$ to as the fundamental solution or the Green function
for the Signalling Problem  of the (linear) cable equation in Eq.(\ref{linear-scaled}),  
whereas to $\H_s$ to as the step response.
As known, the Green function is used in the time convolution integral to represent the solution corresponding to any given input $g(T)$ as follows
  \begin{equation}\label{linsol-general}
 V_m(X,T) =\int_0^T g(T-T')\, \G_s(X,T')\, dT'\,.
\end{equation}
The spatial variance associated to this model is known to evolve linearly in time.

If we consider an impulse or a step current injected at some point $X$ the problem is subjected
to the following boundary conditions, specifically
\begin{equation}
 I=I_0\delta(T)=\frac{-1}{r_i\lambda}\frac{\partial V_m(X,T)}{\partial X}\,,
\end{equation}
or
\begin{equation}
 I=I_0\theta(T)=\frac{-1}{r_i\lambda}\frac{\partial V_m(X,T)}{\partial X}\,.
\end{equation}

We consider the adimensional current $I=I_0 r_i\lambda$ and put it to unity for convenience. Appling the impulse in $X=0$ the LT reduces to
\begin{equation}
 \widetilde{V}_m(X,s)=\frac{1}{\sqrt{s+1}}\e^{-\sqrt{s+1}X}\,,
\end{equation}
the Green function and the step response function (when a step current is applied in $X=0$) reads, respectively,
\begin{equation}
 \G_{m}(X,T)=\frac{1}{\sqrt{\pi T}}\e^{-(\frac{X^2}{4T}+T)}\,,
\end{equation}
and
\begin{equation}
 \H_{m}(X,T) = \int_0^T \frac{1}{\sqrt{\pi T'}}\e^{-(\frac{X^2}{4T'}+T')}\, dT'\,,
\end{equation}

We emphasize that in this standard case the Green function $\G_m(X,T)$
is equal to the Green function for the Cauchy problem $\G_c(X,T)$, for an infinite cable up to constant coefficients.

The motion of ions along the nerve cells is conditionated by this model, that predicts a mean square displacement of diffusing ions that scales linearly with time.
By the way significative deviations from linear behaviour have been measured by experiments, see e.g. Jacobs et al. (1997) \cite{Jacobs-1997},
 Nimchinsky et al. (2002) \cite{Nimchinsky-2002},
 Santamaria et al. (2006) \cite{Fide-2006} and
 Ionescu et al. (2017) \cite{Ionescu}.
 A relevant medical and biological example is 
 the anomalous subdiffusion in neuronal dendritic spines,
 see Suetsugu and Mehraein (1980) \cite{Motohir-1980},
 Duan (2003) \cite{Duan-2003}.   
Particularly appropriate systems are spiny Purkenje cell dendrites characterized by both spiny and not spiny branches.
Spiny branches are in fact characterized by subdiffusive dynamics, while not spiny branches are not. The spatial variance of a diffusing inert tracer (concentration of)
in spiny branches evolves as a sub-linear power law of time, and the diffusion with smaller values of the power exponent is associated
to higher spine density \cite{Fide-2006},
as spines behave as a trap for the diffusing molecules.

 Anomalous subdiffusion can be modelled in several ways introducing  
 some fractional component into the classical cable model. 
The fractional cable model developed in this section is defined by replacing the first order time derivative in Eq.(\ref{linear-scaled}) 
 with a  fractional derivative of order $\alpha\in(0,1)$ of Caputo type see e.g. Gorenflo and Mainardi (1997) \cite{GorMai-CISM97} and Podlubny (1999) \cite{Podlubny-BOOK99}.

 \begin{equation}\label{frac}
 \frac{\partial^2 V_m(X,T)}{\partial X^2}-\frac{\partial^\alpha V_m(X,T)}{\partial T^\alpha}- V_m(X,T)=0\,.
\end{equation}

The solutions of the most relevant boundary problems (Signalling Problem, Cauchy Problem, Second Kind Boundary Problem)
are explicitly calculated in integral form containing  Wright functions. 
Thanks to 
the variability of the parameter $\alpha$, the corresponding  solutions are expected  to  better describe the qualitative 
behaviour of the membrane potential observed in experiments respect to the standard case $\alpha=1$.

From a mathematical point of view this model is a simple extension to fractional behaviour of the Neuronal Cable Model
and it turns to be in some special cases (vanishing initial conditions) equivalent to the equation developed by Henry et al. (2008) \cite{Henry2008}:
\begin{equation}
\frac{\partial V_m(X,T)}{\partial T}= \, _{RL}D_{0,T}^{1-\alpha}\frac{\partial^2 V_m(X,T)}{\partial X^2} -\, _{RL}D_{0,T}^{1-\alpha} (V_m - i_er_m). 
\end{equation}
which can be derived from a modified Nernst-Planck equation, with diffusion constant replaced by fractional derivatives of Riemann-Liouville type.
Other studies consider similar approaches,
see e.g. 
Langlands et al. (2009) \cite{T-2009},
Langlands et al. (2011) \cite{Langlands-2011},
Liu et al. (2011)  \cite{Liu-2011},
Moaddy et al. \cite{Moaddy},
often concentrating on the Initial Value Problem (Cauchy Problem). 
The introduction of a Caputo time derivative to model the voltage response in neurons has already been proposed by Teka et al. (2014) \cite{Teka} to model spiking adaptation for
a homogeneous membrane patch, where the space derivatives vanish, named fractional leaky integrate-and-fire model:
\begin{equation}
\frac{\partial^\alpha V(T)}{\partial T^\alpha}=-(V(T)-V_L)+I_{inj} \,, 
\end{equation}
where an external injected current $I_{inj}$ is included.

Beside the apparent simplicity 
our approach allows to reproduce at least qualitatively the main characteristics observed in experiments \cite{Nimchinsky-2002},
\cite{Jacobs-1997},\cite{Duan-2003},\cite{Motohir-1980}\cite{Fide-2006}. 
Fractional calculus is often used to catch in a parsimonious mathematical description some underlying complex behaviour.
We remind that Caputo's fractional derivative is a non-local operator and for this reason, as pointed out in \cite{Teka},
it could be also introduced to explain behaviours like multiple timescale dynamics  
and memory effects, related to the complexity of the medium. 
Further generalizations
of this model as well the introduction of external injected current and a
more directly relation with ions motion, could be analysed in future, to refine
the biological relevance of the model.

\section*{Solution of the Signalling Problem via Laplace Transform}

The solution of the Signalling Problem has been derived via LT 
by Vitali and Mainardi \cite{Vitali}, however
the inversion of the LT solution for  Eq. (\ref{frac}) requires special efforts because of
the term $V_m(X, T)$. 

When this term is not present, the resulting equation is the well known time fractional diffusion equation:
 \begin{equation}\label{frac-standard}
 \frac{\partial^2 V^*_m(X,T)}{\partial X^2}-\frac{\partial^\alpha V^*_m(X,T)}{\partial T^\alpha}=0\,.
\end{equation}
for which the solutions of the corresponding Cauchy and Signalling Problems have
been derived in the 1990's by Mainardi in terms of two auxiliary Wright functions (of the second type)
\cite{Mainardi-CSF96, Mainardi-CISM97}. Specifically for the Signalling Problem,
the general solution there provided in integral convolution form reads
\begin{equation}\label{sol-frac-standard-W}
 \begin{array}{ll}
 V^*_m(X,T)
 &=\int_0^T g(T-T')\, \G^*_{\alpha, s}(X, T')\, dT'\,,\\ \\
\G^*_{\alpha, s}(X,T)& = 
\frac{1}{T} W_{-{\alpha/2},0}\left(-X/T^{\alpha/2}\right) \,,
\end{array}
\end{equation}
where 
 $\G^*_{\alpha, s}(X,T) $ denotes  the Green function of the Signalling Problem of the fractional time diffusion equation (Eq.(\ref{frac-standard})) and  
$W_{-{\alpha/2},0}(\cdot)$ is a particular case of the transcendental function   known as Wright function
\begin{equation} \label{Wright-function}
 W_{\lambda, \mu }(z) := \sum_{n=0}^{\infty}
 \frac{ z^n}{  n!\, \Gamma[\lambda  n + \mu ]}\,,\quad
 \lambda>-1, \; \mu\ge 0\,.
 \end{equation}
 This  function, entire in the complex plane, 
 is discussed extensively in the Appendix F of Mainardi's book (2010) \cite{Mainardi-BOOK10} where the interested reader can find the 
 following relevant LT pairs, rigorously derived by Stankovic (1970) \cite{Stankovic-WRIGHT70}:
 \begin{equation}\label{Stankovic}
  t^{\mu-1} \, W_{-\nu, \mu} \left(-x/t^\nu\right)
\, \div \, s^{-\mu} \,\exp{\left(-x s^\nu\right)} \ 
  \,,
  \quad 0\le \nu<1\,, \; \mu>0\,.
  \end{equation}
  Here we have adopted an obvious notation to denote the juxtaposition of a locally integrable 
  function of time $t$ with its LT in $s$ with $x$ a positive parameter.
 It is worth to  recall  the distinction of the Wright functions in first type 
 ($\lambda \ge 0$)
  and second type ($-1<\lambda\le 0$)  and, among the latter ones, the relevance of the two auxiliary functions introduced in \cite{Mainardi-CSF96}:
 \begin{equation} \label{F-M}
 F_\nu(z) =  W_{-\nu, 0}(-z)\,, 
  \quad M_\nu(z) =    W_{-\nu, 1-\nu}(-z)\,, \quad 0<\nu<1\,,
  \end{equation}
  inter-related  as $F_\nu(z) = \nu z M_\nu(z) $.
 Indeed the relevance of both the Wright functions
has been outlined by several authors  in diffusion and stochastic processes. 
Particular attention is due  to the   $M$-Wright function (also referred to as the Mainardi function in the book of Podlubny (1999) \cite{Podlubny-BOOK99}) 
  that, since  for $\nu=1/2$ reduces to $\exp{(-z^2/4)}/ \sqrt{\pi}$,
is considered  a suitable generalization of the Gaussian 
density, see Pagnini (2013) \cite{Pagnini-FCAA13} and references therein.

Then the Green function for the Signalling Problem of the time fractional diffusion equation (Eq.(\ref{frac-standard})) can be written in the original form provided by Mainardi (1996) \cite{Mainardi-CSF96} 
 as
 \begin{equation}\label{sol-frac-standard-F-M}
 \G^*_{\alpha, s}(X,T) = 
\frac{1}{T} F_{\alpha/2}\left(X/T^{\alpha/2}\right) =
\frac{\alpha}{2}\, \frac{X}{ T^{\alpha/2+1}}\,
M_{\alpha/2}\left(X/T^{\alpha/2}\right)\,,
\end{equation}
where the superscript $*$  is added  to distinguish the time fractional diffusion equation from our  fractional cable equation, both depending on the order 
$\alpha\in (0,1)$.
   
   Applying the LT to Eq.(\ref{frac}  with the  boundary conditions
required by the Signalling Problem, that is    $V_m(X,0^+)=0$, $V_m(0,T)=g(T)$,  we have:
\begin{equation}
 (s^\alpha+1)\widetilde{V_m}(X,s)- \frac{\partial^2 \widetilde{V_m}(X,s)}{\partial X^2}=0,
\end{equation}
which is a second order equation in the variable $X$ with solution:
\begin{equation}\label{LT-cable}
 \widetilde{V_m}(X,s)= \widetilde{g}(s) e^{ - \sqrt{(s^\alpha + 1)} \cdot X} .
\end{equation}
Because of the shift constant in the square root of the LT 
in Eq.(\ref{LT-cable}),  the inversion is no longer  straightforward with the Wright functions as it is in the time fractional diffusion equation (Eq.(\ref{frac-standard})). 
Consequently,  we 
have overcome this difficulty recurring to the application of the Efros  theorem, see e.g. the book by Graf (2004)  \cite{Graf}  that generalizes the well known convolution theorem for LTs. 
For sake of  convenience let us  hereafter recall this theorem, usually not so well-known  in the literature. 
 The Efros theorem states that if we can write a LT 
 $\widetilde{f}(s)$ as: 
\begin{equation}
\widetilde{f}(s) = \phi(s) \cdot \widetilde{F}(\psi(s)) ,
\end{equation}
where the function $\widetilde{F}(s)$ has a known inverse LT $F(T)$,
the inverse LT can be written in the form: 
\begin{equation}
 f(T)= \int_0^\infty F(\tau)G(\tau,T)d\tau
\end{equation}
where:
\begin{equation}
G(\tau;T)\div \widetilde{G}(\tau,s)= \phi(s) e^{-\tau \psi(s)}
\end{equation}
In Eq.(\ref{LT-cable}), LT solution
of our Signalling Problem,  
  we thus have:
\begin{equation}
 \phi(s)=\widetilde{g}(s),\quad \psi(s)=s^\alpha\,,
\end{equation}
and
\begin{equation}
\widetilde{F}(s)|_{X}  =\e^{-X\sqrt{s+1}}\,.
\end{equation}
Then, having $\widetilde{G}(\tau,s)=
\widetilde{g}(s)\, \e^{-\tau s^\alpha}$, thanks to the standard convolution theorem of LTs, we obtain:
\begin{equation}
 G(\tau,T)=\int_0^T \frac{g(T-T')}{T'}W_{-\alpha,0}(-\tau/T'^\alpha)dT'
\end{equation}
where $W_{-\alpha,0}$ is the  F-Wright function, and 
\begin{equation}
F(T)|_{X}= \frac{X}{\sqrt{4\pi T^3}}e^{-(\frac{X^2}{4T}+T)}
\end{equation}
is the solution in Eq.(\ref{linsol})
 of the standard  cable equation (Eq.(\ref{linear-scaled})).

Then, the general solution for the Signalling Problem can be written in terms of known functions:
\begin{eqnarray}\label{general solution}
 V_m(X,T)&&=\int_0^\infty \frac{X}{\sqrt{4\pi \tau^3}}\,\e^{-(\frac{X^2}{4\tau}+\tau)}
 \left[\int_0^T \frac{g(T-{T'})}{{T'}}W_{-\alpha,0}(-\tau/{T'}^\alpha)d{T'}\right]\, d\tau
 \nonumber\\
 &&=\int_0^T g(T-{T'})\left [\int_0^\infty \frac{X}{\sqrt{4\pi \tau^3}}
 {\rm e}^{-(\frac{X^2}{4\tau}+\tau)}
\frac{1}{T'} F_{\alpha}(\frac{\tau}{T'^\alpha}) \,d\tau \right]\,d{T'}\,.
\end{eqnarray}

\begin{figure}[h!]\centering
 \includegraphics[scale=0.33]{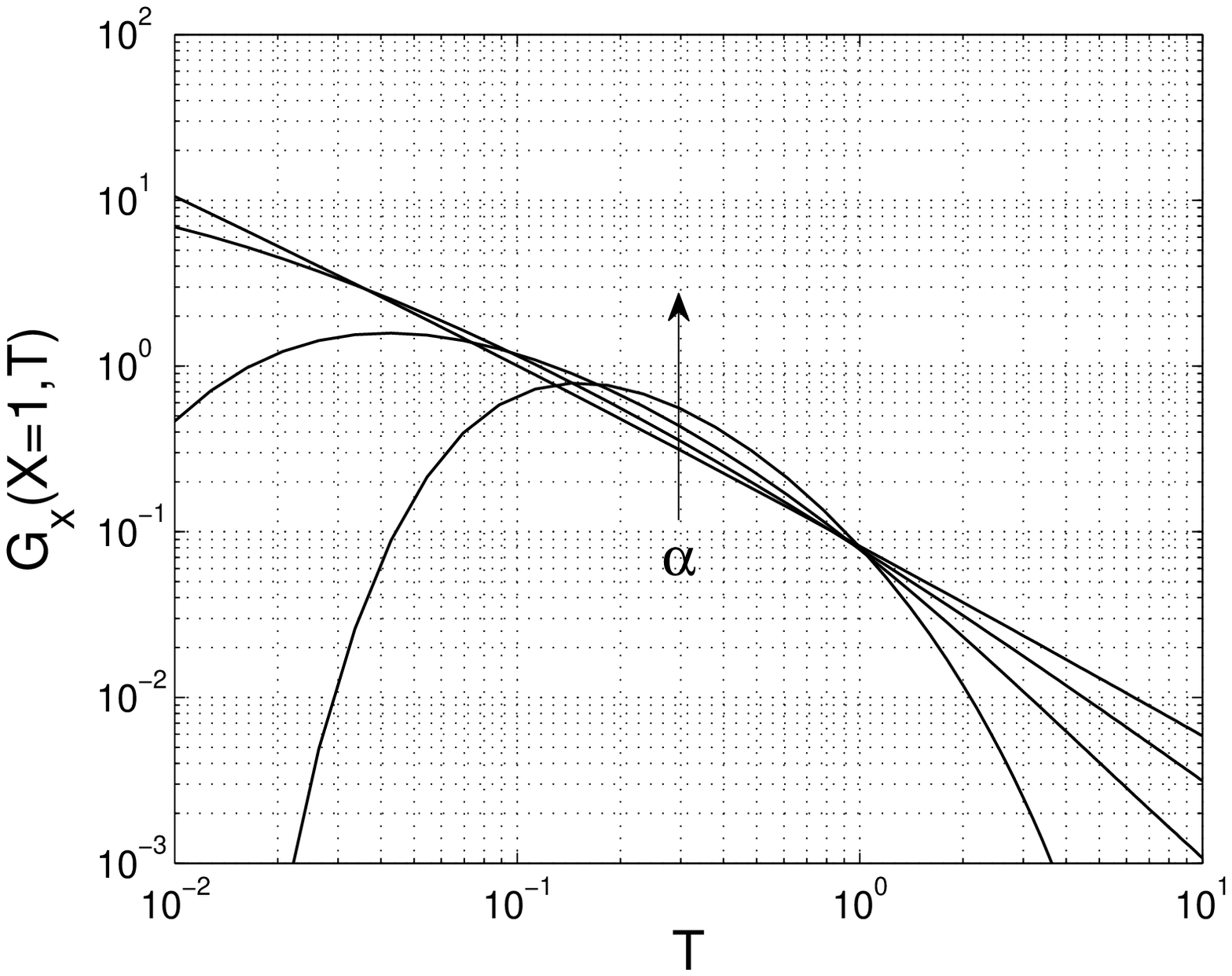} 
 \includegraphics[scale=0.33]{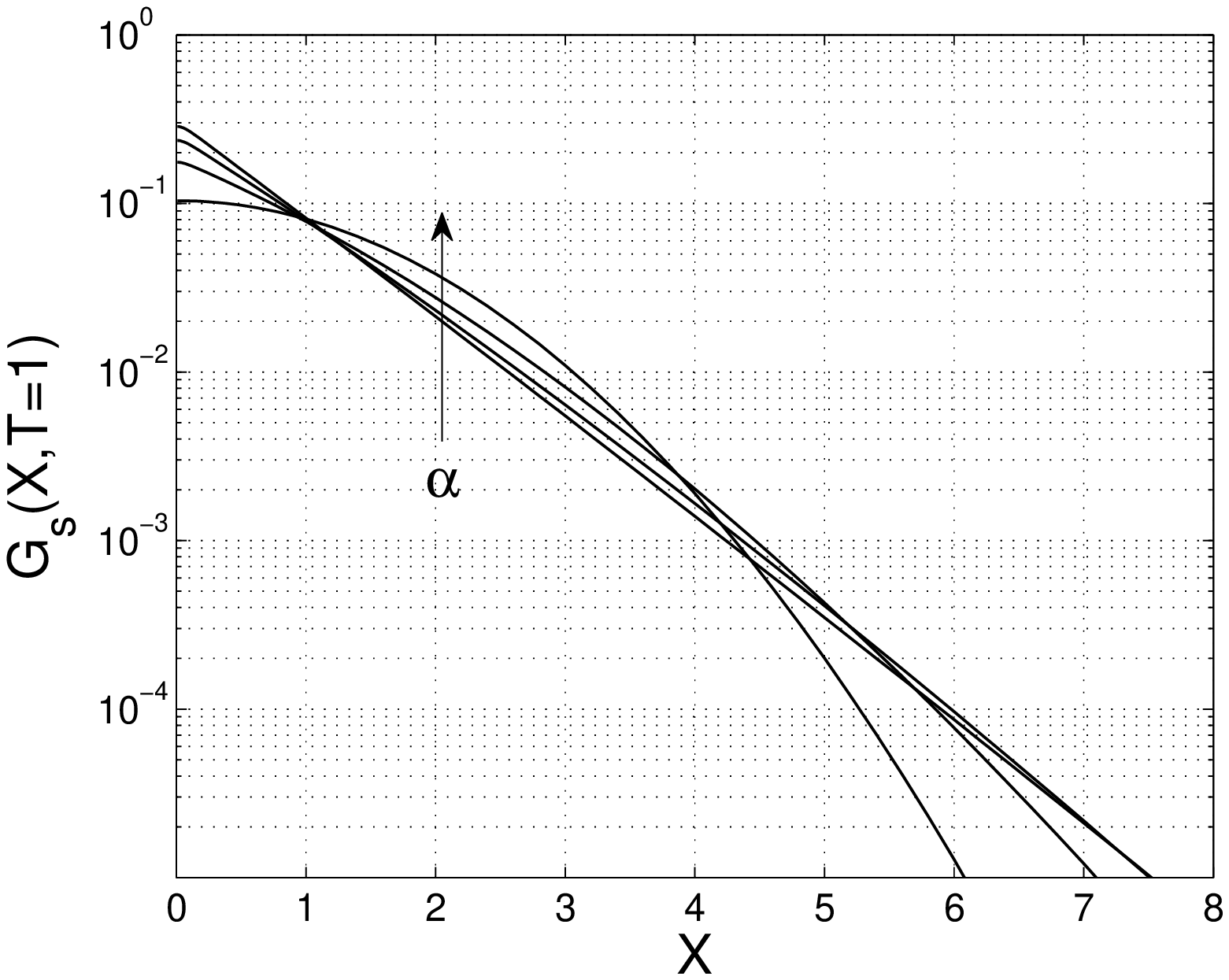}
\caption[Green function for the Signalling Problem]{Green function for Signalling Problem is calculated and plotted for $X=1$ as function of time $T$ (left panel) and
for $T=1$ as function of $X$ (right panel).
Several values of parameter $\alpha$ are compared: $0.25,0.5,0.75,1.$.}
\label{signal1}
 \end{figure}

Substituting $g(T)=\delta(T)$ in the general solution 
in Eq.(\ref{general solution})
we obtain the Green function for the fractional model (Eq.(\ref{frac})), shown in Fig.(\ref{signal1}):
\begin{eqnarray}
V_m(X,T) := 
 \G_{\alpha,s}(X,T)&&=\int_0^\infty \G_s(X,\tau)
\frac{1}{T} F_{\alpha}(\frac{\tau}{T^\alpha}) d\tau\nonumber\\
&&=\int_0^\infty \G_s(X,\tau) \G^*_{2\alpha, s}(\tau,T) d\tau
\end{eqnarray}

\begin{figure}[h!]\centering
  \includegraphics[scale=0.33]{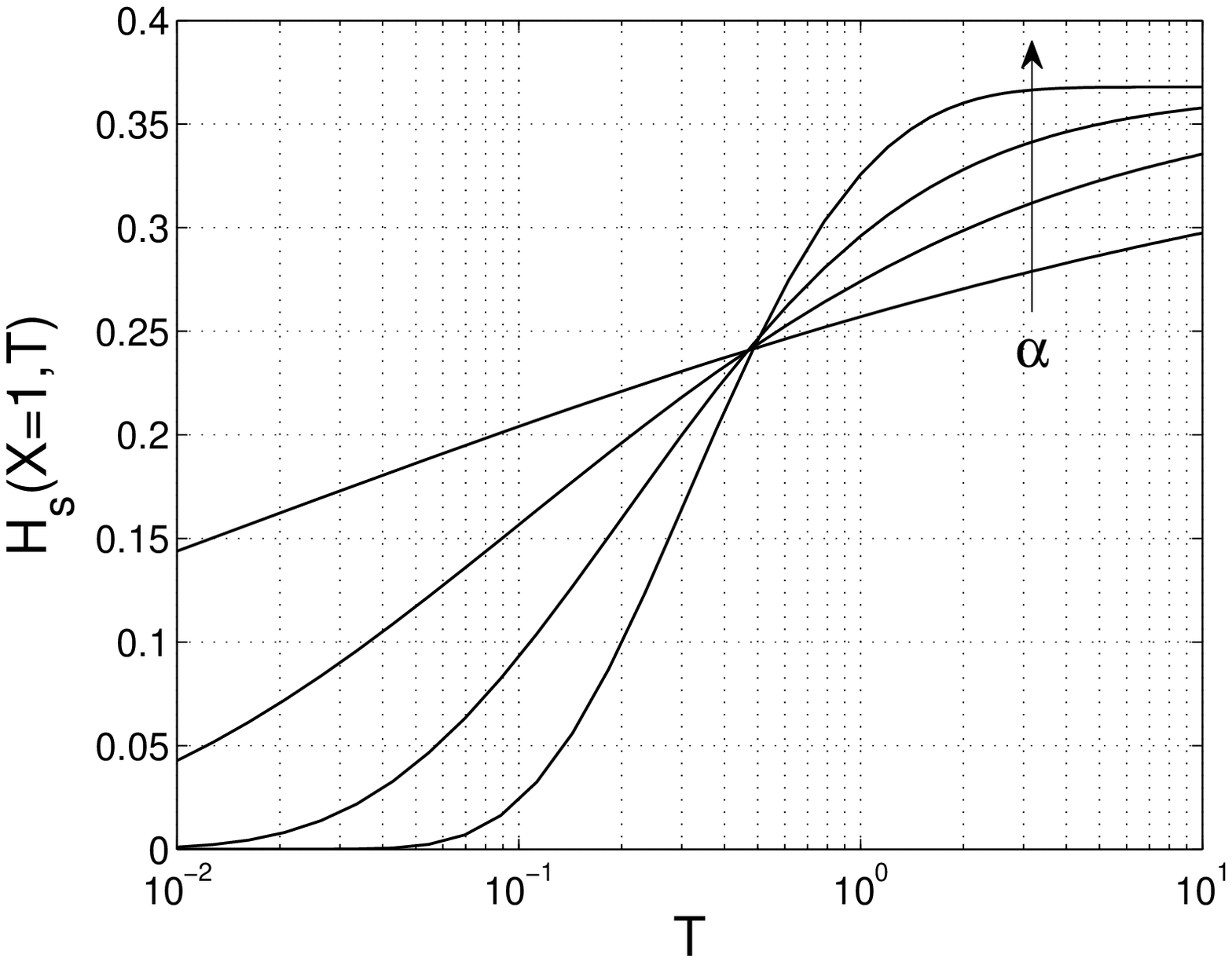} 
 \includegraphics[scale=0.33]{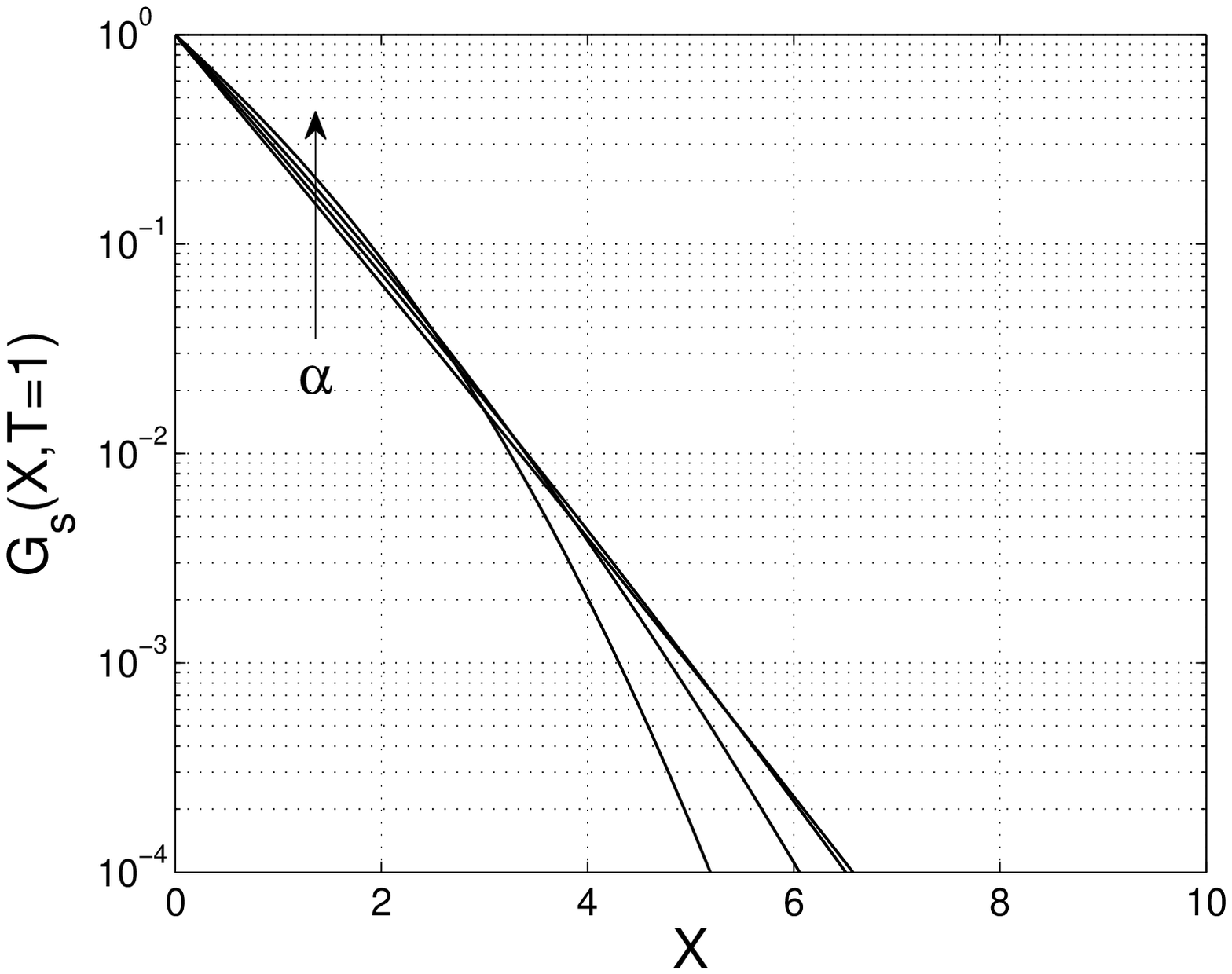} 
\caption[Step Response function for the Signalling Problem]{Step response function for Signalling Problem is 
calculated and plotted for $X=1$ as function of time $T$ (left panel) and
for $T=1$ as function of $X$ (right panel).
Several values of parameter $\alpha$ are compared: $0.25,0.5,0.75,1.$.}
\label{signal2}
 \end{figure}

When $g(T)=\theta(T)$
 we obtain the step response of our fractional cable equation :
\begin{eqnarray}
 V_m(X,T) := 
 \H_{\alpha,s}(X,T) 
 &&=\int_0^\infty  \G_s(X,\tau) 
 \left[\int_0^T \G^*_{2\alpha, s}(\tau,T')d{T'}\right] \, d\tau\\
  &&=\int_0^\infty  \G_s(X,\tau) 
  \H^*_{2\alpha, s}(\tau,T) \, d\tau\,,
\end{eqnarray}
where $ \H^*_{2\alpha, s}(\tau,T) $ is the step response function for the time fractional diffusion equation.
After some manipulations including the change of variable $z=\tau/{T'}^\alpha$ and
integrating by parts after using  the recurrence relation
of Wright functions:
\begin{equation}
\frac{dW_{\lambda,\mu}(z)}{dz}=W_{\lambda,\lambda+\mu}(z)
\end{equation}
and the relation between the auxiliary functions:
$ F_\nu(z)=\nu z \,M_\nu(z)$
we may rewrite the step-response solution as:
\begin{eqnarray}\label{teta}
  V_m(X,T):= 
 \H_{\alpha,s}(X,T) 
  &&= \int_0^\infty \H_s(X,\tau) \cdot\frac{1}{T^\alpha}M_\alpha(\frac{\tau}{T^\alpha})d\tau\\
  &&=\int_0^\infty \H_s(X,\tau) \G^*_{2\alpha, c}(\tau,T)d\tau
\end{eqnarray}
where $ \H_{\alpha,s}(X,T) $ is the step response function for the standard cable model and $ \G^*_{2\alpha, c}(\tau,T)$
is the fundamental solution of the time fractional diffusion equation for the Cauchy Problem. The same expression can easier be derived 
by direct application of the Efros theorem and is plotted in Fig.\ref{signal2}.

\section*{The Green function for the Cauchy Problem}
 
Consider an infinite cable with boundary conditions $V_m(\pm \infty,T)=0$ 
and initial condition $V_m(X,0)=f(X)$. The general solution of the Cauchy problem is related to the Green function $\G_{\alpha,c}(X,T)$ through the
following relation:
\begin{equation}
 V_m(X,T)=\int_{-\infty}^{+\infty}f(x-\xi)\G_{\alpha,c}(\xi,T)d\xi\,.
\end{equation}  
$\G_{\alpha,c}(X,T)$ can be derived via Laplace Transform: 
\begin{equation}
 (s^\alpha+1)\widetilde\G_{\alpha,c}(X,s)- \frac{\partial^2 \widetilde\G_{\alpha,c}}{\partial X^2}=\delta(X)s^{\alpha-1}\,,
\end{equation}
boundary conditions imposes:
\begin{equation}
 \widetilde\G_{\alpha,c}(X,s) = \begin{cases} c_1(s)e^{-X\sqrt{s^\alpha+1}}, & \mbox{if } X>0 \\ c_2(s)e^{+X\sqrt{s^\alpha+1}}, & \mbox{if } X<0 \end{cases}
\end{equation}

Imposing $\widetilde\G_{\alpha,c}(0^-,s)=\widetilde\G_{\alpha,c}(0^+,s)$ leads to $c_1(s)=c_2(s)$.
Integrating Eq.(\ref{frac}) over $X$ from $0^-$ to $0^+$ we have:
\begin{equation}
 \frac{\partial  \widetilde\G_{\alpha,c}(0^+,s) }{\partial X}-\frac{\partial  \widetilde\G_{\alpha,c}(0^-,s) }{\partial X}=-s^{\alpha-1}
\end{equation}
the coefficients result:
\begin{equation}
 c_1(s)=c_2(s)=\frac{1}{2s^{1-\alpha}\sqrt{s^\alpha+1}}
\end{equation}

the resulting LT of the Green function reads:
\begin{equation}
 \widetilde\G_{\alpha,c}(X,s) = \frac{1}{2s^{1-\alpha}\sqrt{s^\alpha+1}} e^{-X\sqrt{s^\alpha+1}}
\end{equation}
 The inversion can be easily performed for $X>0$, thanks again to the Efros theorem, and extended by symmetry respect to the $X$-axes for $X<0$.
 
Let's consider $\phi(s)=\frac{1}{s^{1-\alpha}}$, $\psi(s)=s^\alpha$, following the theorem we may set $G(\tau,s)=\frac{1}{s^{1-\alpha}}e^{-\tau s^\alpha}$ and 
$F(X,s)= \frac{1}{2\sqrt{s+1}} e^{-X\sqrt{s+1}}$, that have known inverse LT:
\begin{equation}
 F(X,T)=\frac{1}{\sqrt{4\pi T}}e^{-\left(\frac{X^2}{4T}+T\right)}
\end{equation}
and 
\begin{equation}
 G(\tau,T)=\frac{1}{T^\alpha}W_{-\alpha,1-\alpha}(-\tau/T^\alpha)=\frac{1}{T^\alpha}M_\alpha(\tau/T^\alpha)
\end{equation}

\begin{figure}[h!]\centering
  \includegraphics[scale=0.33]{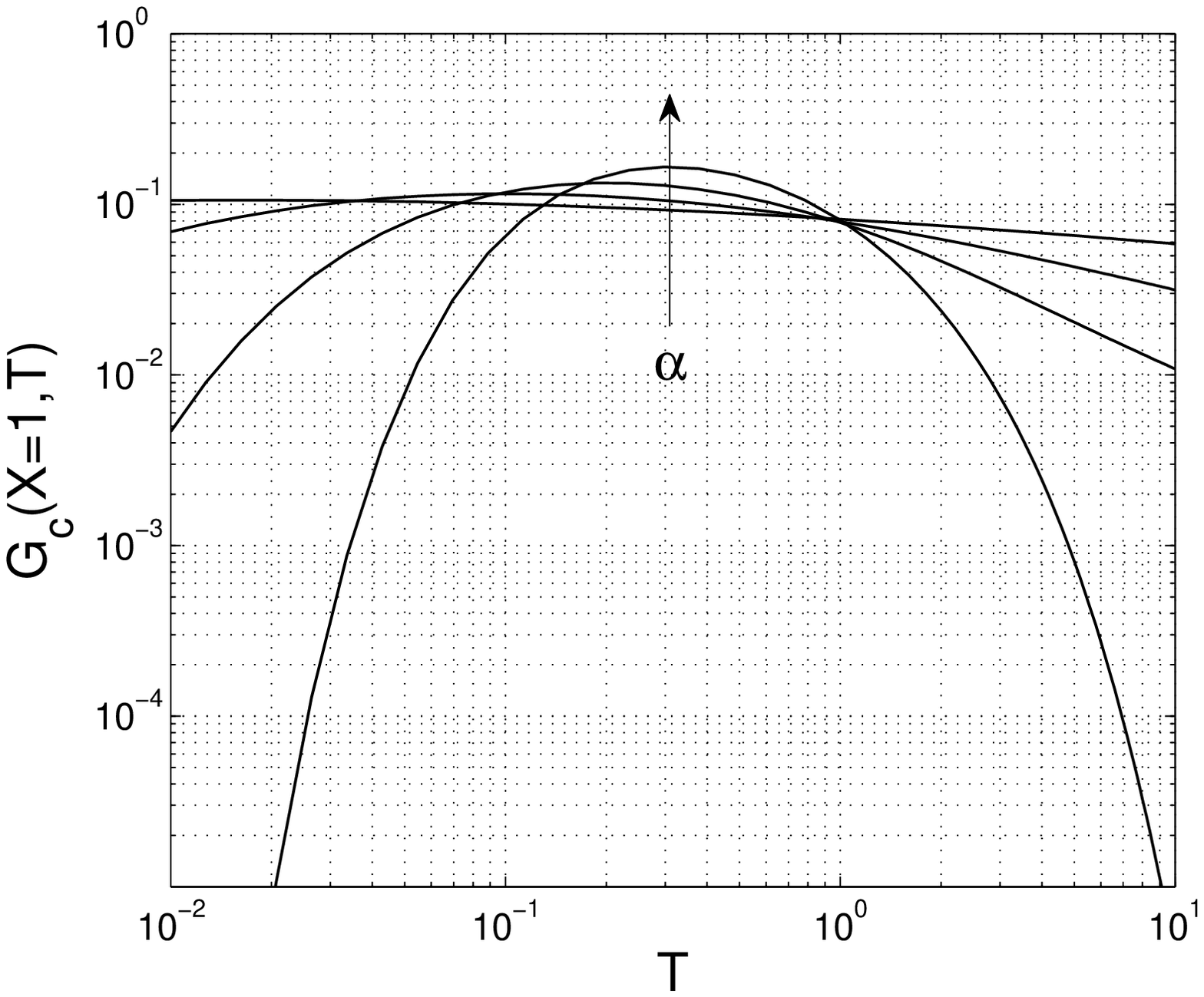} 
 \includegraphics[scale=0.33]{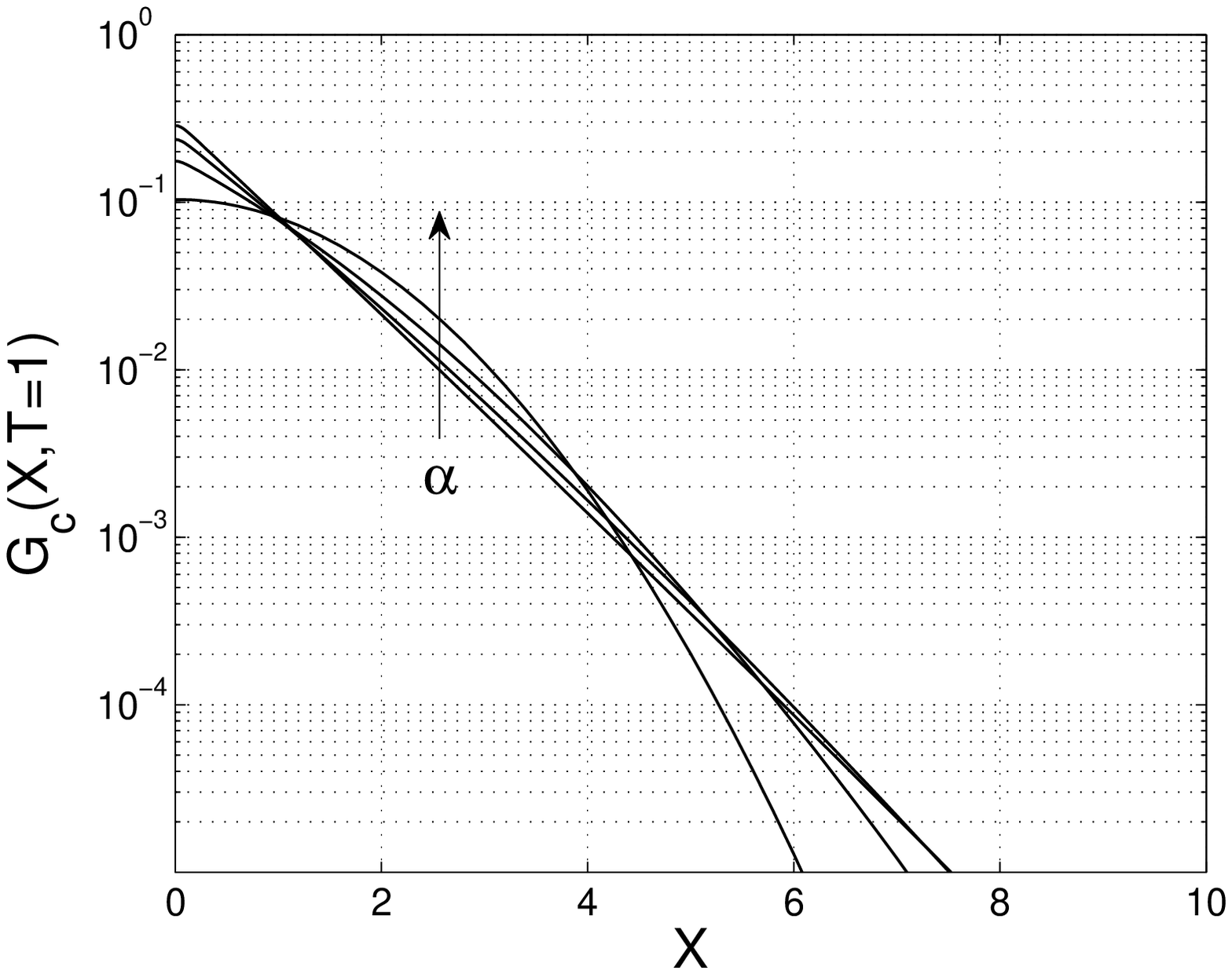} 
\caption[Green function for the Cauchy Problem]{Green function for Cauchy Problem is calculated and plotted for $X=1$ as function of time $T$ (left panel) and
for $T=1$ as function of $X$ (right panel).
Several values of parameter $\alpha$ are compared: $0.25,0.5,0.75,1.$.}
\label{cauchy}
 \end{figure}

The inverse LT for the Green function is plotted in Fig.\ref{cauchy} and reads:
\begin{equation}
\begin{split}
 \G_{\alpha,c}(X,T)&=\int_0^\infty \frac{1}{\sqrt{4\pi \tau}}e^{-\left(\frac{X^2}{4\tau}+\tau\right)}\frac{1}{T^\alpha}M_\alpha(\tau/T^\alpha) d\tau\\
 &=\int_0^\infty \G_{c}(X,\tau) \G^*_{2\alpha, c}(\tau,T) d\tau
\end{split}
 \end{equation}

\section*{Response to injected current}

An interesting biological problem is to consider an injected current in the system.
Transmembrane potential is related to the transmembrane current through the relation $-I=\frac{\partial^2 V_m(X.T)}{\partial X^2}$, where
the minus sign is due to the direction of the current, in this case flowing inside the cell.
Let's consider a singular point 
injected current in $X=0$, it takes the form $I(X,T)=I_0 \delta(X)f(T)$. Integrating from $0^-$ to $0^+$ we obtain the relation
\begin{equation}
 -I_0f(T)=\frac{\partial V_m(X,T)}{\partial X}\big |_{X=0^+}-\frac{\partial V_m(X,T)}{\partial X}\big |_{X=0^-}
\end{equation}

We recall the LT for the semi-infinite cable for an initially undisturbed cable:
\begin{equation}
 \widetilde V_m(X,s) = \widetilde V_m(0,s)e^{-X\sqrt{s^\alpha+1}}\,.
\end{equation}
 At the boundary condition we have:
 \begin{equation}
 I_0 \widetilde f(s) = - \frac{\partial \widetilde V_m(X,s)}{\partial X}\big |_{X=0^+}\,,
 \end{equation}

if we consider an impulse injection of current in $X=0$ we have $I_0\delta(T)=-\frac{\partial V_m(X,T)}{\partial X}\big |_{X=0^+}$. Applying this condition
to the LT we obtain:

\begin{equation}
 \widetilde V_m(0^+,s)=\frac{I_0}{\sqrt{s^\alpha+1}}
\end{equation}

leading to the following Laplace Transformed solution:  
\begin{equation}
  \widetilde \G_{\alpha,m}(X,s) = \frac{I_0}{\sqrt{s^\alpha+1}}
  e^{-X\sqrt{s^\alpha+1}} 
\end{equation}

\begin{figure}[h!]\centering
  \includegraphics[scale=0.33]{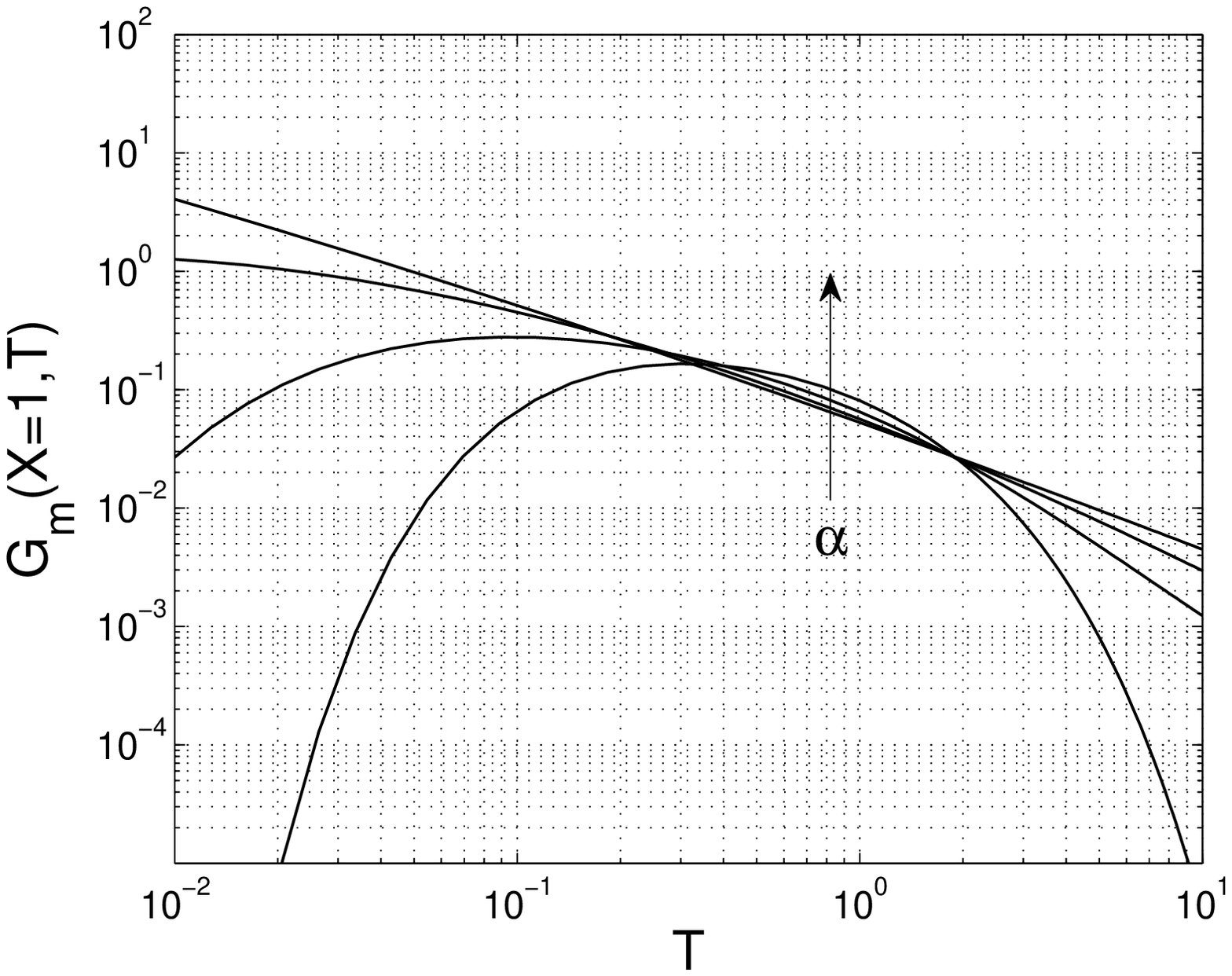} 
 \includegraphics[scale=0.33]{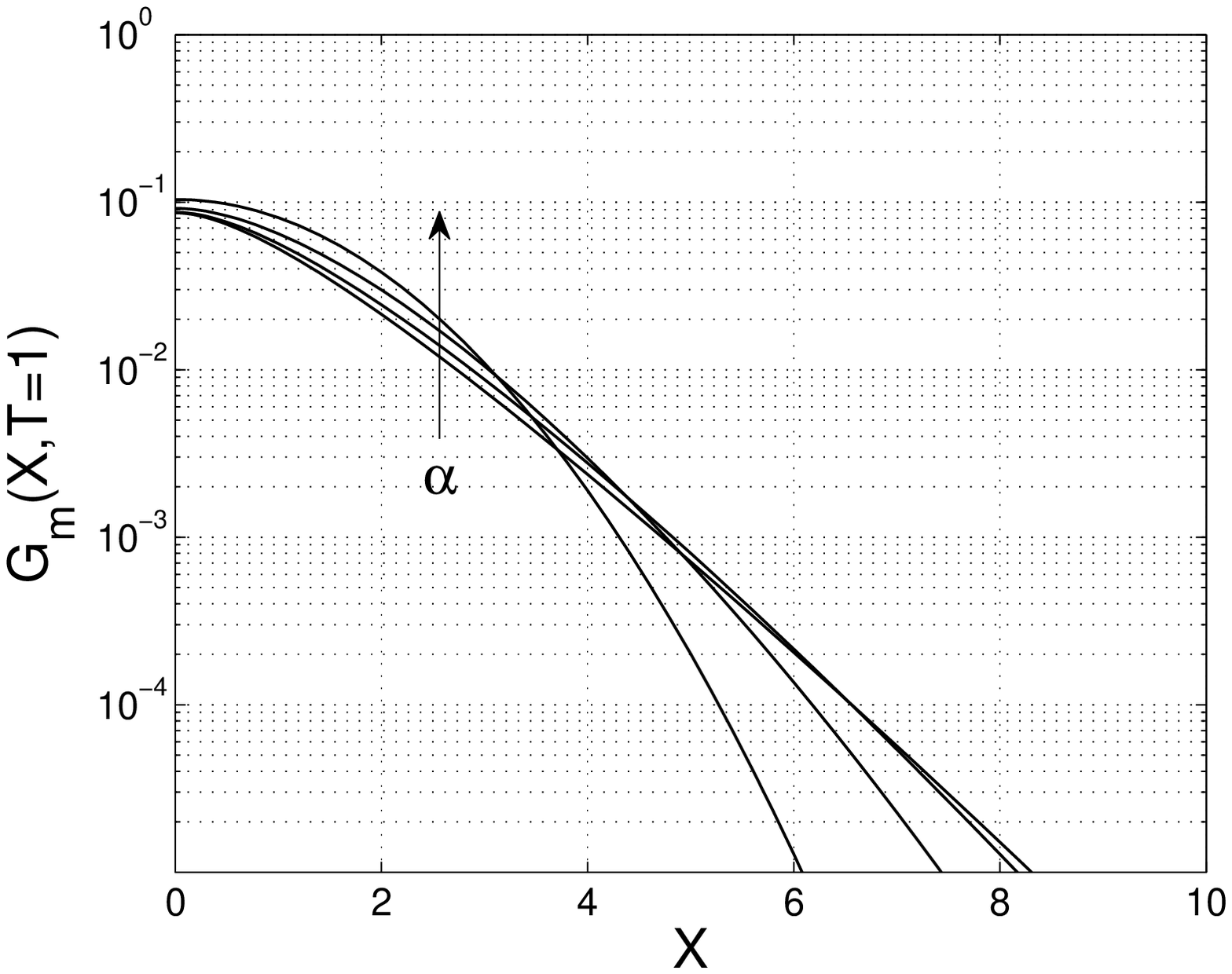} 
\caption[Green function for the Second Kind Boundary Problem]{Green function for 
Second Kind Boundary Problem is calculated and plotted for $X=1$ as function of time $T$ (left panel) and
for $T=1$ as function of $X$ (right panel).
Several values of parameter $\alpha$ are compared: $0.25,0.5,0.75,1.$.}
\label{current}
 \end{figure}

According to the previous derivations it is then straightforward that the inverse LT takes the form:
\begin{equation}
\begin{split}
 \G_{\alpha,m}(X,T)&=\int_0^\infty \frac{I_0}{\sqrt{\pi \tau}}e^{-\left(\frac{X^2}{4\tau}+\tau\right)}\frac{1}{T} W_{-\alpha,0}(-\tau/T^\alpha) d\tau\\
 &=\int_0^\infty \G_{m}(X,\tau)\G^*_{2\alpha, s}(\tau,T) d\tau\,,
\end{split}
 \end{equation}
 represented in Fig.\ref{current}.

For a generic boundary $ I_0 \widetilde f(s)$ we obtain:
\begin{equation}
  \widetilde V_m(X,s) = \frac{I_0f(s)}{\sqrt{s^\alpha+1}}
  e^{-X\sqrt{s^\alpha+1}} 
\end{equation}

The general solution becomes:
\begin{equation}
 V_m(X,T)=\int_0^T f(T-T')\G_{\alpha,m}(X,T')dT'
\end{equation}

The solution is symmetric respect to $X$, the problem can be then extended to the infinite cable introducing a factor $1/2$: 
$\G^\infty_{\alpha,m}(X,T)=\frac{1}{2}\G_{\alpha,m}(X,T)$

The extension to the infinite cable case admits also the following generalization, current injection in $X_0\neq 0$ is equivalent to shift the cable of the same
value $X_0$, then:
\begin{equation}\label{mixed}
 V^\infty_{X_0,m}(X,T)=\int_0^T f(T-T')\G^\infty_{\alpha,m}(X-X_0,T')dT'
\end{equation}

\begin{figure}[h!]\centering
  \includegraphics[scale=0.33]{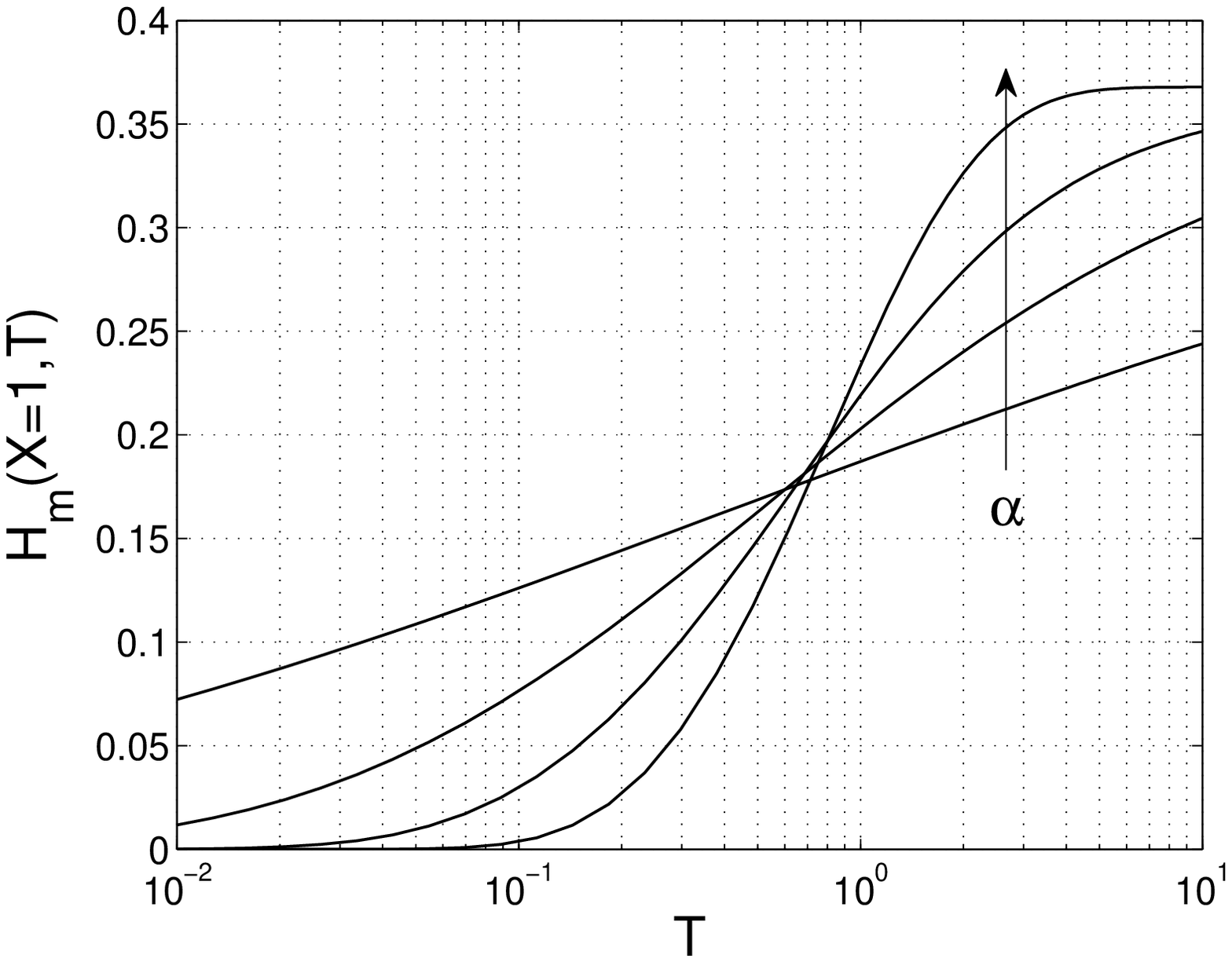} 
 \includegraphics[scale=0.33]{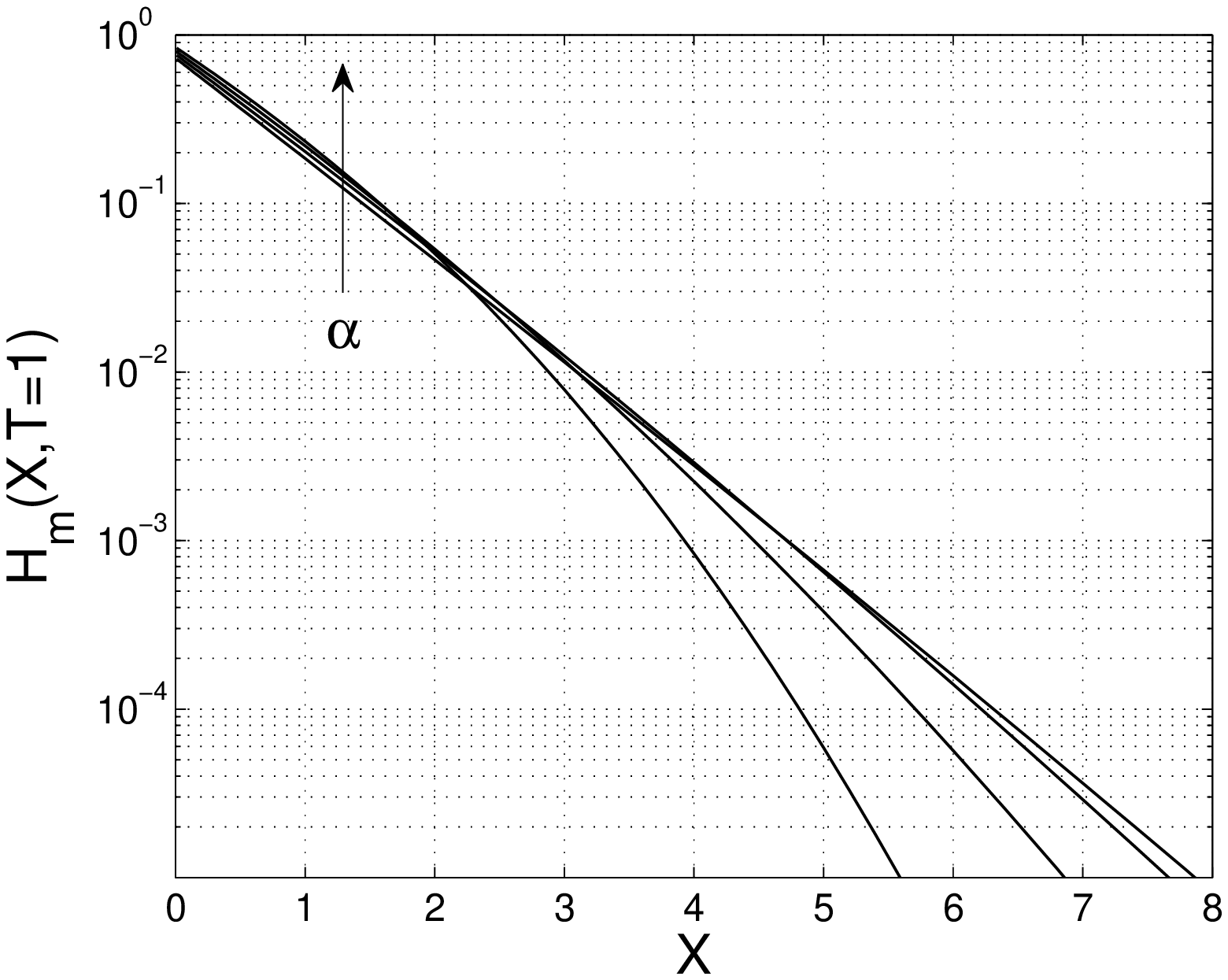} 
\caption[Step Response function for the Second Kind Boundary Problem]{Step response function for 
Second Kind Boundary Problem is calculated and plotted for $X=1$ as function of time $T$ (left panel) and
for $T=1$ as function of $X$ (right panel).
Several values of parameter $\alpha$ are compared: $0.25,0.5,0.75,1.$.}
\label{current2}
 \end{figure}

When the injected current is a step function we obtain the following LT solution:
\begin{equation}
\widetilde \H_{\alpha,m}(X,s)=\frac{I_0}{s\sqrt{s^\alpha+1}}  e^{-X\sqrt{s^\alpha+1}} 
=\frac{I_0}{s^{1-\alpha}s^\alpha\sqrt{s^\alpha+1}} e^{-X\sqrt{s^\alpha+1}} 
\end{equation}
considering $\phi(s)=\frac{1}{s^{1-\alpha}}$, $\psi(s)=s^\alpha$ we have $G(\tau,s)=\frac{1}{s^{1-\alpha}}e^{-\tau s^\alpha}$ and 
$F(X,s)= \frac{1}{s\sqrt{s+1}} e^{-X\sqrt{s+1}}$, that have known inverse LT Eq.(\ref{mixed}) can be simplyfied to:
\begin{equation}
\begin{split}
 \H_{\alpha,m}(X,T)&=\int_0^\infty \H_m(X,\tau)\frac{1}{T^\alpha}M_\alpha(\tau/T^\alpha) d\tau\\
 &=\int_0^\infty  \H_m(X,\tau)\G^*_{2\alpha, c}(\tau,T) d\tau\,,
\end{split}
 \end{equation}
 which is shown in Fig.\ref{current2}.
 
\section*{Conclusions}
The cable model, fractional or linear, is used to describe subthreshold potentials, or passive potentials,
associated to dendritic processes in neurons.
The travelling potential is summed up in the center of the cell, called soma, and an action potential is produced when a threshold is exceeded.
Anomalous regimes of diffusion can then have a deep impact on the communication strength.

Diffusion results more anomalous, i.e. the fractional exponent $\alpha$ decreases,
with increasing spine density \cite{Fide-2006}. Decreasing spine density is characteristic of aging \cite{Jacobs-1997},\cite{Duan-2003}, pathologies
as neurological disorders \cite{Nimchinsky-2002} and Down's sindrome \cite{Motohir-1980}, then subdiffusive regimes are in some sense
associated to a healthy condition. It has been suggested that increasing spine density should serve 
to compensate time delay of postsynaptic potentials along dendrites and to reduce their long time temporal attenuation \cite{Henry2008}.

Looking at our plotted solutions for the fractional cable equation 
when an impulsive potential is applied at the accessible end it can be noted from Fig.\ref{signal1} that peak high decreases more rapidly with decreasing 
$\alpha$ at early times, viceversa is less suppressed at longer times, and the cross over time increases with decreasing $\alpha$.
Looking at the potential versus time it can also be noted that potential functions
associated to lower $\alpha$ last for longer time at 
appreciable intensity and arrive faster at early times with respect to the normal diffusion case ($\alpha=1$).
By the way, when a constant potential is applied  at the accessible end we note from Fig.\ref{signal2}
that the exponential suppression of the potential along the dendrite is reduced for high $X$ values with respect to 
normal diffusion. Instead for small $X$ the potential results just slightly more suppressed in the sub-diffusion process. 
These behaviours can be noticed also for the other cases in Fig.\ref{cauchy} and Fig.\ref{current} - \ref{current2}.

From a mathematical point of view the Efros theorem extends the concept of convolution as an integral form that is consistent with a 
subordination-type integral. 
However such integral form does not necessary connote a subordinated process, as it has been shown in literature for 
the generalized grey Brownian motion (ggBm) by Mura and Pagnini (2008) \cite{Mura-2008} and
in a more extended way by
 Molina-Garcia et al. (2016) \cite{Molina-Garcia-2016}, 
 but could also be interpreted as a consequence of the random nature of the media in which particles are diffusing.
This model can be also read as a generalization of time fractional diffusion processes where mass is not conserved due to leakage.
This approach naturally recover the solution for the time fractional case in the limit in which the leakage is put to zero in the integral forms. 

In conclusion the presented fractional cable model satisfies the main biological features of the dendritic cell Signalling Problem.
With respect to models solved as Cauchy problem, our approach could include specific
time dependent boundary conditions, which will allow to reconstruct with accuracy 
the expected signal at the soma if the model will result capable to predict real data behaviour. 
Furthermore the 
solutions can be computed directly, i.e.
calculating the integral associated, as well as by Laplace Transform inversion,
see e.g. Abate and Ward (2006) \cite{Abate} without any remarkable issue.
Further generalizations of the model as the inclusion of external currents in the equation and comparison with data will be investigated in future works.

\section*{Aknowledgements}
The work of 
F. M.  has been carried out in the framework of the activities of the
 National Group of Mathematical Physics (INdAM-GNFM).
All the authors acknowledge support by the Italian Ministry of Education and the 
Interdepartmental Center  "Luigi Galvani" for integrated studies of Bioinformatics,
Biophysics and Biocomplexity of the University of Bologna.
The authors would also like to thank PhD. G. Pagnini (BCAM, Bilbao, Spain) for valuable comments and discussion and the anonymous referees 
for their constructive remarks and suggestions that helped to improve the manuscript.



\begin{thebibliography}{10}

\bibitem{Abate}
Abate, J. and Ward, W. (2006).
\newblock A unified framework for numerically inverting Laplace Transforms.
\newblock {\em INFORMS Journal on Computing}, {\bf 18}(4), pp. 408--421.


\bibitem{Duan-2003}
Duan, H. (2003).
\newblock Age-related dendritic and spine changes in corticocortically
  projecting neurons in {M}acaque monkeys.
\newblock {\em Cerebral Cortex}, {\bf 13}(9), pp. 950--961.

\bibitem{GorMai-CISM97}
Gorenflo, R. and Mainardi, F. (1997).
\newblock  Fractional Calculus: Integral and Differential Equations of
  Fractional Order, in:
in: A. Carpinteri and F. Mainardi (Editors),
\newblock {\it Fractals and Fractional Calculus in  Continuum Mechanics},
Springer Verlag, Wien and New York (1997),
 pp. 223--276.
[E-print http://arxiv.org/abs/0805.3823]. 

\bibitem{Graf}
Graf, U. (2004).
\newblock {\em Applied Laplace Transforms and z-Transforms for Scientists and
  Engineers}.
\newblock Published by Birkhäuser Basel, ISBN: 978-3-0348-9593-4

\bibitem{Henry2008}
Henry, B., Langlands, T., and Wearne, S. (2008).
\newblock Fractional cable models for spiny neuronal dendrities.
\newblock {\em Physical Review Letters}, {\bf 100}:128103, pp. 1--3.


\bibitem{Ionescu}
Ionescu, C., Lopes, A., Copot, D., Machado, J.A.T. and Bates, J.H.T., (2017).
\newblock The role of fractional calculus in modelling biological phenomena: A review.
\newblock {\em Communications in Nonlinear Science and Numerical Simulation}, {\bf 51}, pp. 141--159.

\bibitem{Jacobs-1997}
Jacobs, B., Driscoll, L., and Schall, M. (1997).
\newblock Life-span dendritic and spine changes in areas 10 and 18 of human
  cortex: A quantitative golgi study.
\newblock {\em The Journal of Comparative Neurology}, {\bf 386}(4), pp. 661--680.

\bibitem{Johnston}
Johnston, D. and Miao-Sin Wu, S. (1994).
\newblock {\em Foundations of Cellular Neurophysiology}.
\newblock Bradford Books. The MIT Press, 1 edition, Cambridge (US).




\bibitem{T-2009}
Langlands, T., Henry, B.~I., and Wearne, S. (2009).
\newblock Fractional cable equation models for anomalous electrodiffusion in
  nerve cells: infinite domain solutions.
\newblock {\em Journal of Mathematical Biology}, {\bf 59}, pp. 761--808


\bibitem{Langlands-2011}
Langlands, T., Henry, B., and Wearne, S. (2011).
\newblock Fractional cable equation models for anomalous electrodiffusion in
  nerve cells: Finite domain solutions.
\newblock {\em SIAM Journal on Applied Mathematics}, {\bf 71}(4), pp. 1168--1203.


\bibitem{Liu-2011}
Liu, F., Yang, Q., and Turner, I. (2011).
\newblock Two new implicit numerical methods for the fractional cable equation.
\newblock {\em Journal of Computational and Nonlinear Dynamics}, 
{\bf 6}(1), pp. 1--7.

\bibitem{Magin}
Magin, R. (2006).
\newblock {\em Fractional Calculus in Bioengineering}.
\newblock Begell House Publishers, Connecticut, US.

\bibitem{Mainardi-CSF96}
Mainardi, F. (1996).
\newblock Fractional relaxation-oscillation and fractional diffusion-wave
  phenomena.
\newblock {\em Chaos, Solitons and Fractals}, {\bf 7}, 
pp. 1461--1477.

\bibitem{Mainardi-CISM97}
Mainardi, F. (1997).
\newblock  Fractional Calculus: Some Basic problems in Continuum and
  Statistical Mechanics, 
in: A. Carpinteri and F. Mainardi (Editors),
\newblock {\it Fractals and Fractional Calculus in  Continuum Mechanics},
Springer Verlag, Wien and New York (1997), pp. 291--348.
[E-print http://arxiv.org/abs/1201.0863] .

\bibitem{Mainardi-BOOK10}
Mainardi, F. (2010).
\newblock {\it  Fractional Calculus and Waves in Linear Viscoelasticity} - Appendix F.
\newblock Imperial College Press. 1st edition. London.

\bibitem{Moaddy}
Moaddy, K., Radwan, A.~G., Salama, K.~N., Momani, S., and Hashim, I. (2012).
\newblock The fractional-order modeling and synchronization of electrically
  coupled neuron systems.
\newblock {\em Computers and Mathematics with Applications}, {\bf 64}(10), pp. 3329--3339.

\bibitem{Molina-Garcia-2016}
Molina-Garcia, D., Pham,  T.M., Paradisi, P., Manzo C. and Pagnini, G. (2016).
\newblock Fractional kinetics emerging from ergodicity breaking in random
  media.
\newblock {\em Physical Review E}, {\bf 94}, 052147.

\bibitem{Mura-2008}
Mura, A. and  Pagnini, G. (2008).
\newblock Characterizations and simulations of a class of stochastic processes
  to model anomalous diffusion.
\newblock {\em Journal of Physics A Mathematical and Theoretical}, {\bf 41}, 285003 (22 pages).
[E-print https://arxiv.org/abs/0801.4879]

\bibitem{Nimchinsky-2002}
Nimchinsky, E.~A., Sabatini, B.~L., and Svoboda, K. (2002).
\newblock Sructure and function of dendritic spines.
\newblock {\em Annual Review of Physiology}, {\bf 64}, pp. 313--353.

\bibitem{Pagnini-FCAA13}
Pagnini, G. (2013).
\newblock The {M}-{W}right function as a generalization of the {G}aussian
  density for fractional diffusion processes.
\newblock {\em Fractional Calculus and Applied Analysis}, {\bf 16}(2), pp. 436--453.

\bibitem{Podlubny-BOOK99}
Podlubny, I. (1999).
\newblock {\it Fractional Differential Equations}.
\newblock Mathematics in Science and Engineering 198. Academic Press, San
  Diego, 1st edition.

\bibitem{Qian}
Qian, N. and Sejnowski, T. (1989).
\newblock An electro-diffusion model for computing membrane potentials and ionic concentrations in branching dendrites, spines and axons, 
\newblock {\em Biological Cybernetics}, {\bf 62}, pp. 1--15 
  
\bibitem{Fide-2006}
Santamaria, F., Wils, S., Schutter, E.~D., and Augustine, G.~J. (2006).
\newblock Anomalous diffusion in purkinje cell dendrites caused by spines.
\newblock {\em Neuron}, {\bf 52}(4), pp. 635--648.

\bibitem{Stankovic-WRIGHT70}
Stankovi\`c, B. (1970).
\newblock On the function of {E}.{M}. {W}right.
\newblock {\em Publ. de l'InstitutMath\`ematique, Beograd, Nouvelle S\`er.}, {\bf 10}, pp. 113--124.

\bibitem{Motohir-1980}
Suetsugu, M. and Mehraein, P. (1980).
\newblock Spine distribution along the apical dendrites of the pyramidal
  neurons in {D}own's syndrome.
\newblock {\em Acta Neuropathologica}, {\bf 50}(3), pp. 207--10.

\bibitem{Teka}
Teka, W. and Marinov, T.M. and Santamaria, F. (2014).
\newblock Neuronal spike timing adaptation described with a fractional leaky integrate-and-fire model.
\newblock {\em PLOS Computational Biology}, {\bf 10} (3), e1003526.

\bibitem{Tukwell}
Tuckwell, H.C.  (1988).
\newblock {\it Introduction to Theoretical Neurobiology - Vol.1: Linear Cable Theory and Dendritic Structure}. 
\newblock  Cambridge University Press, Cambridge.

\bibitem{Vitali}
Vitali, S. and Mainardi, F. (2017).
\newblock Fractional cable model for signal conduction in spiny neuronal
  dendrites.
\newblock {\em AIP Conference Proceedings}, Accepted.

\bibitem{Weiss}
Weiss, T.F. (1996).
\newblock {\it Cellular Biophysics}, Vol. 2: Electrical Properties.
\newblock  Bradford Book. 


\end{thebibliography}
\end{document}